\documentclass[sigconf, authorversion]{acmart}

\usepackage{amsmath}
\usepackage{tabularx}					
\usepackage[flushleft]{threeparttable}	
\usepackage{booktabs}

\usepackage{caption}
\usepackage{multirow}
\usepackage{graphicx}
\usepackage{array}
\usepackage{color}
\usepackage[noend]{algorithmic}
\usepackage{algorithm}
\usepackage{setspace}
\usepackage{pifont}
\usepackage{bbm}
\usepackage{subfigure}

\definecolor{orange}{rgb}{1, .36, .08}

\newcommand{\fakeparagraphnospace}[1]{\noindent\textbf{#1.}}

\setcopyright{acmlicensed}

\usepackage{hyperref}

\usepackage{titlesec}
\titlespacing*{\section}{0pt}{4pt}{1pt}

\begin{document}

\title{
Late Breaking Results: On the One-Key Premise of Logic Locking
} 

\author{
Yinghua Hu \quad Hari Cherupalli \quad Mike Borza \quad Deepak Sherlekar \\ [-2ex]
\small Synopsys, Inc., Sunnyvale, CA, USA \\ \{yinghuah, harich, mborza, dsherlek\}@synopsys.com
}

\copyrightyear{2024}
\acmYear{2024}
\setcopyright{rightsretained}\acmConference[DAC '24]{61st ACM/IEEE Design Automation Conference}{June 23--27, 2024}{San Francisco, CA, USA}
\acmBooktitle{61st ACM/IEEE Design Automation Conference (DAC '24), June 23--27, 2024, San Francisco, CA, USA}
\acmDOI{10.1145/3649329.3663504}
\acmISBN{979-8-4007-0601-1/24/06}

\begin{abstract}
The evaluation of logic locking methods has long been predicated on an implicit assumption that only the correct key can unveil the true functionality of a protected circuit.
Consequently, a locking technique is deemed secure if it resists a good array of attacks aimed at finding this correct key. 
This paper challenges this one-key premise by introducing a more efficient attack methodology, focused not on identifying that one correct key, but on finding multiple, potentially incorrect keys that can collectively produce correct functionality from the protected circuit.
The tasks of finding these keys can be parallelized, which is well suited for multi-core computing environments. 
Empirical results show our attack achieves a runtime reduction of up to $99.6\%$ compared to the conventional attack that tries to find a single correct key.
\end{abstract}


\pagenumbering{gobble}
\maketitle

\vspace{-1mm}
\section{Introduction}\label{sec:intro}
The past two decades have witnessed rapid development of logic locking, a promising hardware intellectual property (IP) protection method to thwart various threats, e.g., IP piracy and hardware Trojan insertion, performed by rogue agents on the semiconductor supply chain~\cite{hu2021risk,cryptoeprint:2022/260}. 
Logic locking techniques are designed to secure a design's netlist by incorporating extra programmable logic components that are regulated through an additional set of input ports called key ports. 
If a wrong key value is applied to these key ports, the altered netlist will produce an incorrect function.
This mechanism enables designers to safely send the locked netlist to a foundry without the risk of IP theft in the supply chain. 
Evaluating the effectiveness of logic locking has led to 
many efficient attacks, e.g.,~\cite{hu2023security,chen2023unraveling}, aimed at discovering the correct unlocking key. Traditionally, a locking method is deemed ``secure'' if it withstands a wide array of these attacks.
However, this conventional wisdom hinges on a critical assumption, i.e., the one-key premise, which posits that obtaining the correct key is essential for accessing the original design's functionality. Surprisingly, this fundamental assumption has received scant attention in literature, raising questions about its validity and the potential implications for the security of logic locking. 

This paper scrutinizes the conventional ``one-key'' premise 
and introduces a novel attack methodology based on 
the Boolean satisfiability (SAT) attack methodology~\cite{subramanyan2015evaluating}. 
Unlike traditional approaches that focus on identifying the single correct key, our method seeks to identify multiple incorrect keys that, when used together, can effectively unlock the protected design as if the correct key were applied. 
Furthermore, this proposed attack can be formulated as a set of independent sub-tasks, which significantly shortens the attack time in a multi-core environment—a capability readily exploitable by resource-rich adversaries in the supply chain. 
Our empirical evidence shows that the proposed attack can achieve up to $99.6\%$ runtime reduction, 
demonstrating that, under the ``multi-key'' assumption, many existing logic locking techniques can be much less secure than they appear.

\section{Preliminaries \& Background}\label{sec:background}
We denote the original circuit as $\mathcal{C}$, with the sets of input and output ports labeled as $I$ and $O$, respectively. Consequently, the function of $\mathcal{C}$, denoted by $f$, is defined as $f: \mathbb{B}^{|I|} \rightarrow \mathbb{B}^{|O|}$. For the locked circuit, represented as $\mathcal{C}_l$, its function $f_l$ incorporates an additional set of key ports, $K$, introduced during the locking process, and is given by $f_l: \mathbb{B}^{|I|} \times \mathbb{B}^{|K|} \rightarrow \mathbb{B}^{|O|}$. The correct unlocking key, $k^*$, satisfies the condition $f_l(i, k^*) = f(i)$, $\forall i \in \mathbb{B}^{|I|}$, distinguishing it from incorrect keys that produce erroneous outputs for at least one input pattern.

\fakeparagraphnospace{Satisfiability Attack}
The SAT attack~\cite{subramanyan2015evaluating} is among the most effective attacks against logic locking. The attack assumes two primary resources: (1) the reverse-engineered locked design netlist $\mathcal{C}_l$ from the GDSII file used by the foundry, and (2) the original design's golden input-output pairs, obtainable through querying a commercially available chip. This attack initiates by constructing a miter circuit from duplicate instances of $\mathcal{C}_l$, which is then analyzed by a SAT solver. The solver's objective is to identify a specific input pattern, known as a Distinguishing Input Pattern (DIP), that exposes output discrepancies between the two instances of $\mathcal{C}_l$ under different keys. Each identified DIP, along with its corresponding output obtained from the ``oracle'', i.e., the working chip, helps 
eliminate incorrect keys by refining the miter circuit.
The process iterates
until no further DIPs can be found, indicating the successful elimination of all incorrect keys from the search space.
A key in the remaining search space will be returned as the correct key. 
Since its creation, the SAT attack has become a standard method to evaluate new logic locking techniques. The count of DIPs needed to uncover the correct key, symbolized as $\#DIP$, serves as a preliminary gauge of a locking method's security. 
Additionally, certain locking techniques, like LUT-based insertion~\cite{chowdhury2021enhancing}, complicate the SAT attack by increasing the miter instance's complexity after each iteration, thus leading to longer SAT solving time over successive iterations. 
Consequently, the duration of the attack is often employed as a more precise measure of a locking technique's resilience.

\fakeparagraphnospace{Countermeasures against the SAT Attack}
Numerous strategies have been developed to counter the SAT attack, broadly falling into two categories: (1) augmenting the required number of DIPs, and (2) enhancing the complexity of the miter to identify a DIP.
\emph{SARLock}~\cite{yasin2016sarlock} is an example of the first category, adding a ``point function'' component to the original design. 
While SARLock corrupts only a small portion of the design function, 
$\#DIP$
exponentially grows with the key size.
Conversely, to markedly elevate the complexity of the miter, various computationally demanding logic elements, such as Look-up Tables (LUTs)~\cite{chowdhury2021enhancing} and eFPGAs~\cite{mohan2021hardware}, are embedded into the original design netlist. 
While attacking these enhanced schemes may not necessitate a large $\#DIP$, 
the resolution of each DIP demands a considerable amount of time.
In this paper, we benchmark our proposed attack against SARLock and LUT-based insertion. 

\section{Attack Methodology}
\begin{figure}[t]
\centering
\subfigure[]{
 \resizebox{0.6\columnwidth}{!}{
 \raisebox{65pt}{
 \vspace{-5mm}
\begin{tabular}{|cc|cccccccc|}
\hline
\multicolumn{2}{|l|}{\multirow{2}{*}{\textbf{Error Distribution}}} & \multicolumn{8}{c|}{\textbf{Key}}                                                                                                                                                                         \\ \cline{3-10} 
\multicolumn{2}{|l|}{}                                    & \multicolumn{1}{c|}{\textbf{000}} & \multicolumn{1}{c|}{\textbf{001}} & \multicolumn{1}{c|}{\textbf{010}} & \multicolumn{1}{c|}{\textbf{011}} & \multicolumn{1}{c|}{\textbf{100}} & \multicolumn{1}{c|}{\textbf{101}} & \multicolumn{1}{c|}{\textbf{110}} & \textbf{111} \\ \hline
\multicolumn{1}{|c|}{\multirow{8}{*}{\textbf{Input}}}     & \textbf{000}    & \multicolumn{1}{c|}{\textcolor{red}{\ding{55}}}    & \multicolumn{1}{c|}{\checkmark}    & \multicolumn{1}{c|}{\checkmark}    & \multicolumn{1}{c|}{\checkmark}    & \multicolumn{1}{c|}{\checkmark}    & \multicolumn{1}{c|}{\checkmark}    & \multicolumn{1}{c|}{\checkmark}    &   \checkmark  \\ \cline{2-10} 
\multicolumn{1}{|c|}{}                           & \textbf{001}    & \multicolumn{1}{c|}{\checkmark}    & \multicolumn{1}{c|}{\textcolor{red}{\ding{55}}}    & \multicolumn{1}{c|}{\checkmark}    & \multicolumn{1}{c|}{\checkmark}    & \multicolumn{1}{c|}{\checkmark}    & \multicolumn{1}{c|}{\checkmark}    & \multicolumn{1}{c|}{\checkmark}    &   \checkmark  \\ \cline{2-10} 
\multicolumn{1}{|c|}{}                           & \textbf{010}    & \multicolumn{1}{c|}{\checkmark}    & \multicolumn{1}{c|}{\checkmark}    & \multicolumn{1}{c|}{\textcolor{red}{\ding{55}}}    & \multicolumn{1}{c|}{\checkmark}    & \multicolumn{1}{c|}{\checkmark}    & \multicolumn{1}{c|}{\checkmark}    & \multicolumn{1}{c|}{\checkmark}    &   \checkmark  \\ \cline{2-10} 
\multicolumn{1}{|c|}{}                           & \textbf{011}    & \multicolumn{1}{c|}{\checkmark}    & \multicolumn{1}{c|}{\checkmark}    & \multicolumn{1}{c|}{\checkmark}    & \multicolumn{1}{c|}{\textcolor{red}{\ding{55}}}    & \multicolumn{1}{c|}{\checkmark}    & \multicolumn{1}{c|}{\checkmark}    & \multicolumn{1}{c|}{\checkmark}    &   \checkmark  \\ \cline{2-10} 
\multicolumn{1}{|c|}{}                           & \textbf{100}    & \multicolumn{1}{c|}{\checkmark}    & \multicolumn{1}{c|}{\checkmark}    & \multicolumn{1}{c|}{\checkmark}    & \multicolumn{1}{c|}{\checkmark}    & \multicolumn{1}{c|}{\textcolor{red}{\ding{55}}}    & \multicolumn{1}{c|}{\checkmark}    & \multicolumn{1}{c|}{\checkmark}    &   \checkmark  \\ \cline{2-10} 
\multicolumn{1}{|c|}{}                           & \textbf{101}    & \multicolumn{1}{c|}{\checkmark}    & \multicolumn{1}{c|}{\checkmark}    & \multicolumn{1}{c|}{\checkmark}    & \multicolumn{1}{c|}{\checkmark}    & \multicolumn{1}{c|}{\checkmark}    & \multicolumn{1}{c|}{\checkmark}    & \multicolumn{1}{c|}{\checkmark}    &   \checkmark  \\ \cline{2-10} 
\multicolumn{1}{|c|}{}                           & \textbf{110}    & \multicolumn{1}{c|}{\checkmark}    & \multicolumn{1}{c|}{\checkmark}    & \multicolumn{1}{c|}{\checkmark}    & \multicolumn{1}{c|}{\checkmark}    & \multicolumn{1}{c|}{\checkmark}    & \multicolumn{1}{c|}{\checkmark}    & \multicolumn{1}{c|}{\textcolor{red}{\ding{55}}}    &   \checkmark  \\ \cline{2-10} 
\multicolumn{1}{|c|}{}                           & \textbf{111}    & \multicolumn{1}{c|}{\checkmark}    & \multicolumn{1}{c|}{\checkmark}    & \multicolumn{1}{c|}{\checkmark}    & \multicolumn{1}{c|}{\checkmark}    & \multicolumn{1}{c|}{\checkmark}    & \multicolumn{1}{c|}{\checkmark}    & \multicolumn{1}{c|}{\checkmark}    &  \textcolor{red}{\ding{55}}     \\ \hline
\end{tabular}
\label{tab:error_dist}
}
}
}
\subfigure[]{
\includegraphics[width=0.28\columnwidth]{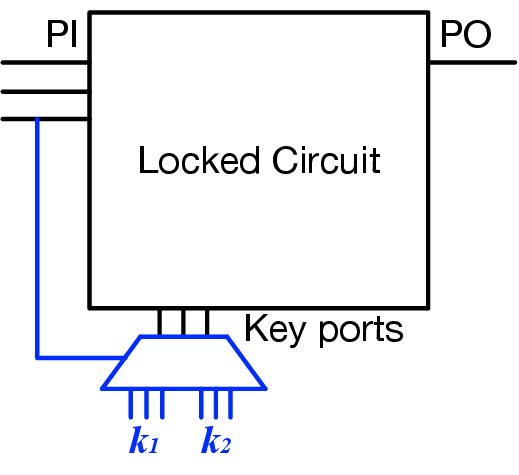}\label{fig:mux_scheme}
}
\vspace{-5mm}
\caption{(a) Error distribution on an example circuit ($|I|=|K|=3$), and (b) the schematic of the modified locked netlist.}
\vspace{-8mm}
\label{fig:two_figs}
\end{figure}

As a foundation of our improved attack, we explore the feasibility of utilizing multiple incorrect keys to access the full design functionality. 
This approach is predicated on the idea that the locked design's function, $f_l$, can be decomposed into several terms, where each term is contingent upon distinct conditions applied to the primary inputs.
For instance, one such condition might involve the most significant bit (MSB) of the input ports being set to either a logical high or low state. We represent the corresponding decomposed function of $f_l$ as follows:
\begin{equation}\label{eq:divide}
    \begin{aligned}
    f_l(i,k) = f_l(i_{[\text{msb}=0]},k) \vee f_l(i_{[\text{msb}=1]},k),
    \end{aligned}
\end{equation}
where each term encapsulates one half of the overall locked function $f_l$.
Similarly to $f_l$, the two sub-functions defined in equation~\eqref{eq:divide} can be unlocked with the correct key. However, it is notable that each sub-function might also be accessible using different incorrect keys. For instance, consider the error distribution for a 3-input circuit locked with SARLock, as depicted in Fig.~\ref{tab:error_dist}. Here, the sole correct key is $101$. Yet, by examining the sub-function for when the MSB is zero (the top four rows), we identify three incorrect keys ($100$, $110$, and $111$) that also successfully unlock this portion of the function. A similar pattern emerges when analyzing the sub-function for the MSB set to one (the bottom four rows). 
Suppose the attacker already possesses two keys, each capable of unlocking a different half of the design's function. Fig.~\ref{fig:mux_scheme} illustrates a schematic of the adjusted locked netlist. 
In this netlist, the two keys serve as input to a multiplexer (MUX) module, which operates based on the same criterion used for function splitting—specifically, the MSB's logic value. This alteration renders the netlist functionally equivalent to the one unlocked with the correct key, underscoring the potential for the collective usage of multiple, potentially incorrect keys to access the full design function.

\begin{algorithm}[t]
\begin{spacing}{0.8}  
 \caption{Proposed Attack Flow}
 \begin{algorithmic}[1]\label{alg:generic_flow}
 \renewcommand{\algorithmicrequire}{\textbf{Input:}}
 \renewcommand{\algorithmicensure}{\textbf{Output:}}
 \small
\REQUIRE Locked netlist $\mathcal{C}_l$, black-box oracle $\mathcal{C}$, splitting effort $N$
\ENSURE  A list of $N$ keys $\mathcal{K}$
\STATE $\mathcal{K} = list()$
\FOR{$i=1$ \TO $2^N$}
    \STATE $b$ = convert\_to\_binary\_and\_pad($i$, $N$)
    \STATE $\mathcal{C}_{temp}$ = generate\_conditional\_netlist($\mathcal{C}_l$, $b$)
    \STATE $k$ = SAT\_attack($\mathcal{C}_{temp}$, $\mathcal{C}$)
    \STATE $\mathcal{K}$.append($k$)
\ENDFOR
\RETURN $\mathcal{K}$
\end{algorithmic} 
\end{spacing}
\end{algorithm}
\setlength{\textfloatsep}{2pt}
Algorithm~\ref{alg:generic_flow} outlines our enhanced SAT attack designed to identify multiple keys, with each key unlocking a specific portion of the locked function. 
The attacker inputs the locked netlist $\mathcal{C}_l$, an oracle $\mathcal{C}$, and a splitting effort $N$, thereby dividing the input space into $2^N$ distinct terms. 
Each term is contingent on setting $N$ primary input ports to the binary sequence $b$ generated in line 3 and is subsequently synthesized to remove any redundant logic (line 4). 
The resulting netlist, $\mathcal{C}_{temp}$, undergoes the traditional SAT attack process (line 5). The key reported by the SAT attack is then added to the key list $\mathcal{K}$ (line 6). 
The algorithm concludes by repeating the aforementioned steps for all the $2^N$ distinct terms. 
Although Algorithm~\ref{alg:generic_flow} is presented as a sequential process, it is worth highlighting that the tasks within each distinct term
are autonomous and can be executed independently of one another. 
By leveraging the capacity for parallel execution, attackers can adjust the splitting effort $N$ in alignment with their computational resources, significantly reducing the total duration of the attack.

\section{Validation Results}

Our enhanced attack is primarily implemented in \textsc{Python}, with \textsc{Mini-SAT} utilized for SAT solving to maintain consistency with the traditional SAT attack method. For the synthesis of the modified locked netlists, we employ the Synopsys Design Compiler and the 45-nm Nangate Open Cell Library~\cite{nangate}. To ensure a fair comparison, we use the ISCAS'85 benchmark circuits~\cite{hansen1999unveiling}, aligning with the datasets reported in both SARLock~\cite{yasin2016sarlock} and LUT-based insertion studies~\cite{chowdhury2021enhancing}. 
All experiments are executed on a server with $16\times$ $2.6$-GHz cores, leading to the range of the splitting effort $N$ to be from $1$ to $4$. 
The selection of which $N$ input ports to apply the splitting condition is determined through a fan-out cone analysis of the netlist's input ports, prioritizing those with the most key-controlled gates in their fan-out cones. This approach is based on the premise that input ports influencing wider key-controlled areas 
can significantly simplify the netlist's logic when held constant, thereby further reducing the complexity of miters used in the SAT solving.

\fakeparagraphnospace{Attacking SARLock}
Since the number of DIPs ($\#DIP$) required to break SARLock is deterministic, we first apply our enhanced attack on SARLock as a flow checker. 
Table~\ref{tab:sarlock_result} shows the results of attacking a benchmark circuit $c7552$, locked with different key sizes. 
\renewcommand{\arraystretch}{0.8} 
\begin{table}[t]
    \caption{$\#DIP$ Results for SARLock-locked $c7552$.} 
    \vspace{-5mm}
    \begin{center}
    \resizebox{0.67\columnwidth}{!}{
    \begin{tabular}{|c|c|c|c|c|c|}
\hline
  & $N=0$ (baseline) & $N=1$ & $N=2$ & $N=3$ & $N=4$ \\ \hline
$|K|=4$  &       14                    &  6   &   3  &   1  &  1   \\ \hline
$|K|=8$  &           255                &  127   &    63 & 31    &  15   \\ \hline
$|K|=12$ &            4095               &  2047   &  1023   &    511 &  255   \\ \hline
\end{tabular}
    }
    \label{tab:sarlock_result}
    \end{center}
    \vspace{-4mm}
\end{table}
When $N>0$, we observe the same $\#DIP$ for all the parallelized tasks. 
It is obvious to see that $\#DIP$ exponentially deceases as $N$ goes up. 

\fakeparagraphnospace{Attacking LUT-based Insertion}
\begin{table}[t]
    \caption{Runtime (seconds) of Attacking LUT-based Insertion.} 
    \vspace{-5mm}
    \begin{center}
    \resizebox{0.9\columnwidth}{!}{
    \begin{tabular}{|c|c|cccc|}
\hline
\multirow{2}{*}{Circuit} & \multirow{2}{*}{Baseline~\cite{subramanyan2015evaluating}} & \multicolumn{4}{c|}{This work}                                                                                \\ \cline{3-6} 
                         &                           & \multicolumn{1}{c|}{Minimum} & \multicolumn{1}{c|}{Mean}    & \multicolumn{1}{c|}{Maximum} & \textbf{Maximum/Baseline} \\ \hline
c880                     & 42311.8                   & \multicolumn{1}{c|}{80.1}    & \multicolumn{1}{c|}{127.1}   & \multicolumn{1}{c|}{169.9}   & \textbf{0.004}            \\ \hline
c1355                    & 26533.5                   & \multicolumn{1}{c|}{84.0}    & \multicolumn{1}{c|}{133.7}   & \multicolumn{1}{c|}{213.2}   & \textbf{0.008}            \\ \hline
c1908                    & 15005.3                   & \multicolumn{1}{c|}{124.6}   & \multicolumn{1}{c|}{152.9}   & \multicolumn{1}{c|}{198.2}   & \textbf{0.013}           \\ \hline
c2670                    & 91403.3                   & \multicolumn{1}{c|}{17763.6} & \multicolumn{1}{c|}{38103.8} & \multicolumn{1}{c|}{57304.0} & \textbf{0.627}            \\ \hline
c3540                    & 158416.0                  & \multicolumn{1}{c|}{123.5}   & \multicolumn{1}{c|}{314.0}   & \multicolumn{1}{c|}{560.3}   & \textbf{0.004}            \\ \hline
c5315                    & 1774.5                    & \multicolumn{1}{c|}{665.8}   & \multicolumn{1}{c|}{2358.9}  & \multicolumn{1}{c|}{5627.4}  & \textbf{3.171}            \\ \hline
c6288                    & 8652.8                    & \multicolumn{1}{c|}{0.172}   & \multicolumn{1}{c|}{102.3}   & \multicolumn{1}{c|}{213.3}   & \textbf{0.025}            \\ \hline
c7552                    & 48815.3                   & \multicolumn{1}{c|}{170.0}   & \multicolumn{1}{c|}{316.8}   & \multicolumn{1}{c|}{544.1}   & \textbf{0.011}           \\ \hline
\end{tabular}
    }
    \label{tab:lut_result}
    \end{center}
\end{table}
Aligned with the locking methodology described in~\cite{chowdhury2021enhancing}, we incorporate a $14$-input $2$-stage LUT module into the benchmark circuits, effectively equating to a key size of $156$. 
To maximize the advantage of our proposed attack, we use a splitting effort of $N=4$. 
Finally, we present in Table~\ref{tab:lut_result} the runtime of the traditional SAT attack and our proposed attack. 
Within our proposed attack, we detail the minimum, maximum, and average of the $16$ parallel tasks. 
Given that our attack's efficiency is determined by the runtime of the most time-intensive sub-task, we report our method's advantage using the ratio of the maximum sub-task runtime to that of the baseline attack. 
Remarkably, except for $c5315$, all other benchmarks exhibit an average runtime reduction of $90.1\%$, with the maximum reduction being as high as $99.6\%$. 
For six out of the eight benchmarks, the maximum-to-baseline runtime ratio falls below $1/16=0.0625$, suggesting significant efficiency improvement even within a single-core computing context.
This result primarily stems from dividing the key searching task into smaller segments, which not only result in smaller SAT instances to solve but also increase the likelihood of more keys capable to unlock the sub-functions. 

\section{Conclusion}
We challenge the traditional ``one-key'' premise in logic locking techniques with an innovative attack strategy capable of efficiently identifying multiple, potentially incorrect keys that, collectively, can unlock the protected design. 
Future works include creating effective defenses to counter the new ``multi-key'' attack scenario. 

\bibliographystyle{ieeetr}
\bibliography{references} 
\end{document}